\documentclass[pre,
   superscriptaddress,
   twocolumn,
   showpacs,
   preprintnumbers,
   byrevtex]{revtex4-1}

\usepackage{amsfonts}
\usepackage{amssymb}
\usepackage{amsmath}

\usepackage{graphicx}
\usepackage{color}
\usepackage{hyperref}

\begin{document}

\preprint{LA-UR-10-02381} 

\title{Band structure of SnTe studied by Photoemission Spectroscopy}

\author{P.B.~Littlewood}
\affiliation{Cavendish Laboratory, Cambridge University, JJ Thomson Ave, Cambridge CB3 0HE, United Kingdom}

\author{B.~Mihaila}
\affiliation{Materials Science and Technology Division, Los Alamos National Laboratory, Los Alamos, New Mexico, USA}
\author{R.K.~Schulze}
\affiliation{Materials Science and Technology Division, Los Alamos National Laboratory, Los Alamos, New Mexico, USA}
\author{D.J.~Safarik}
\affiliation{Materials Science and Technology Division, Los Alamos National Laboratory, Los Alamos, New Mexico, USA}

\author{J.E.~Gubernatis}
\affiliation{Theoretical Division, Los Alamos National Laboratory, Los Alamos, New Mexico, USA}

\author{A.~Bostwick}
\affiliation{Advanced Light Source, Lawrence Berkeley National Laboratory, Berkeley, California, USA}
\author{E.~Rotenberg}
\affiliation{Advanced Light Source, Lawrence Berkeley National Laboratory, Berkeley, California, USA}

\author{C.P.~Opeil}
\affiliation{Department of Physics, Boston College, Chestnut Hill, Massachusetts, USA}

\author{T.~Durakiewicz}
\affiliation{Materials Physics and Applications, Los Alamos National Laboratory, Los Alamos, New Mexico, USA}

\author{J.L.~Smith}
\affiliation{Materials Science and Technology Division, Los Alamos National Laboratory, Los Alamos, New Mexico, USA}
\author{J.C.~Lashley}
\affiliation{Materials Science and Technology Division, Los Alamos National Laboratory, Los Alamos, New Mexico, USA}

\begin{abstract}
We present an angle-resolved photoemission spectroscopy study of the electronic structure of SnTe, and compare the experimental results to {\em ab initio} band structure calculations as well as a simplified tight-binding model of the p-bands. Our study reveals the conjectured complex Fermi surface structure near the L-points showing topological changes in the bands from disconnected pockets, to open tubes, and then to cuboids as the binding energy increases, resolving lingering issues about the electronic structure. 
The chemical potential at the crystal surface is found to be 0.5~eV below the gap, corresponding to a carrier density 
of $p =1.14\times10^{21}$~cm$^{-3}$ or $7.2 \times 10^{-2}$~holes per unit cell.
At a temperature below the cubic-rhombohedral structural transition a small shift in spectral energy of the valance band is found, in agreement with model predictions.
\end{abstract}

\maketitle


The IV-VI compounds, of which SnTe is an exemplar, have long been of interest for two reasons in particular. One is that these pseudo-cubic materials are narrow gap semiconductors or semimetals with excellent thermoelectric and bolometric properties~\cite{thermoelectrics}. The second, and related property is that they are structurally ``soft''. Heats of formation are small, reflecting the weak ionicity; consequently they are easily transformed between glassy and crystalline states. Because the small and reversible structure changes strongly influence the conductivity, these materials are employed in phase-change memory applications (particularly in the mixed alloy series of (IV-VI)-(V$_2$VI$_3$)): the glass state is semiconducting, whereas the crystal is metallic, though there is no detailed microscopic explanation for this effect~\cite{wuttig}.
Below 100~K SnTe transforms in a continuous structural transition from a rocksalt to a rhombohedral structure, produced by a small dimerization within the cubic unit cell along the (111) direction, which is a polar version of the structure of the isoelectronic group~V semimetals As, Sb, and Bi~\cite{cowley}.

As grown, cubic SnTe is typically metallic and p-type due to slight off-stoichiometry. Earlier work~\cite{burke65,tung69,allgaier72,gordyunin74,melvin79} has established that the holes are confined to pockets at the L-points on the surface of the Brillouin zone. These pockets are highly anisotropic and non-parabolic, and the non-parabolicity significantly affects the transport properties even at relatively small hole concentrations of $10^{18}$~cm$^{-3}$. The carrier properties have been modeled to compare with transport, Hall effect, and optical properties to good effect. However, there has hitherto been no direct measurement of the electronic structure, and the electronic states at high doping $\ge 10^{20}$~cm$^{-3}$ have not been firmly established.

In this paper we present a angle-resolved photoemission study of a IV-VI compound. We find that the surface carrier density is substantial, and by tuning in energy below the chemical potential we follow the evolution of the energy surface from pockets, to elongated tubes connecting the L-points, and eventually into the deeper parts of the band structure which are deformed cubes. This is reproduced essentially exactly by {\em ab initio} band structure calculations, but may also be understood quite simply within a tight-binding picture of p-orbitals. At energies further below the Fermi energy, we also observe the expected gap opening corresponding to the structural phase transition, providing a quantitative test of theoretical models of the origin of the structural transition.


SnTe single crystals 
were prepared by fusion of the elements in a Bridgman furnace.
Angle-resolved photoemission 
experiments were carried out at the Advanced Light Source Beamline 7.0.1. using synchrotron radiation ($h\nu=111$eV). 
The sample was cleaved in situ at 20~K to expose a fresh single-crystal (001) surface. 
The photoemission data was compared with results of first-principle band structure calculations performed using the generalized gradient-approximation approach  GGA~\cite{gga} in the full-potential linearized-augmented-plane-wave  method~\cite{wien2k}.

\begin{figure*}[t]
\centering{\includegraphics[width=0.8\linewidth]{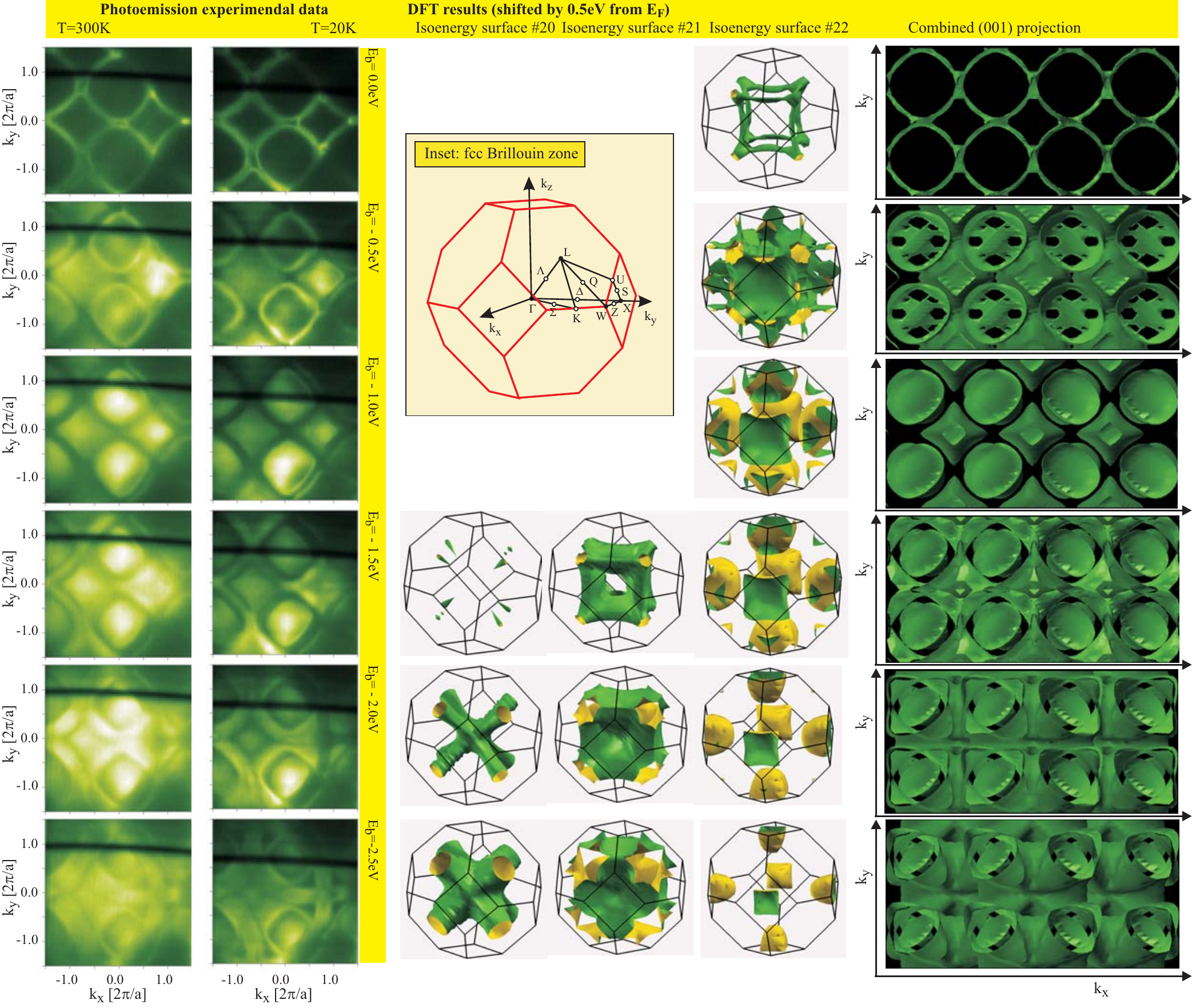}}
\caption{(Color online) Photoemission spectra at T=300~K  and T=20~K  at a sequence of energies.
The right panels show the calculated isoenergy surfaces projected onto the (100) direction, whereas
the 3D isoenergy surfaces for the three relevant bands are shown in the center to clarify the topology. 
Here, the indexing of the calculated isoenergy surfaces corresponds to the electronic band index in the DFT calculation. Band \#22 reaches closest to the Fermi surface.  
The inset illustrates the fcc Brillouin zone.
We note that the dark line at k$_y=1$ is due to shadowing during data collection.}
\label{Fig1}
\end{figure*}


Momentum-space cuts along the (001) direction at a sequence of fixed energies are shown in Fig.~\ref{Fig1}, at room temperature and also at a low temperature well below the structural transition.
It is evident that there is a large open Fermi surface, which is unexpected given that Hall measurements on this sample showed a small carrier density, $3.57\times10^{19}$~cm$^{-3}$ at 77~K, indicative of the expected small hole pockets at the L-points. The symmetry is consistent with the expected cleave along a (100) face, and there is no evidence for surface reconstruction. Because this is a material with three-dimensional (3D) character, these data, of course, represent projections into a 2D plane. For further analysis we turn to a comparison with band structure. We shall first present {\em ab initio} numerical results and then show how a simple tight-binding model can reproduce the main points.

\begin{figure}[t]
\centering{\includegraphics[width=0.9\columnwidth]{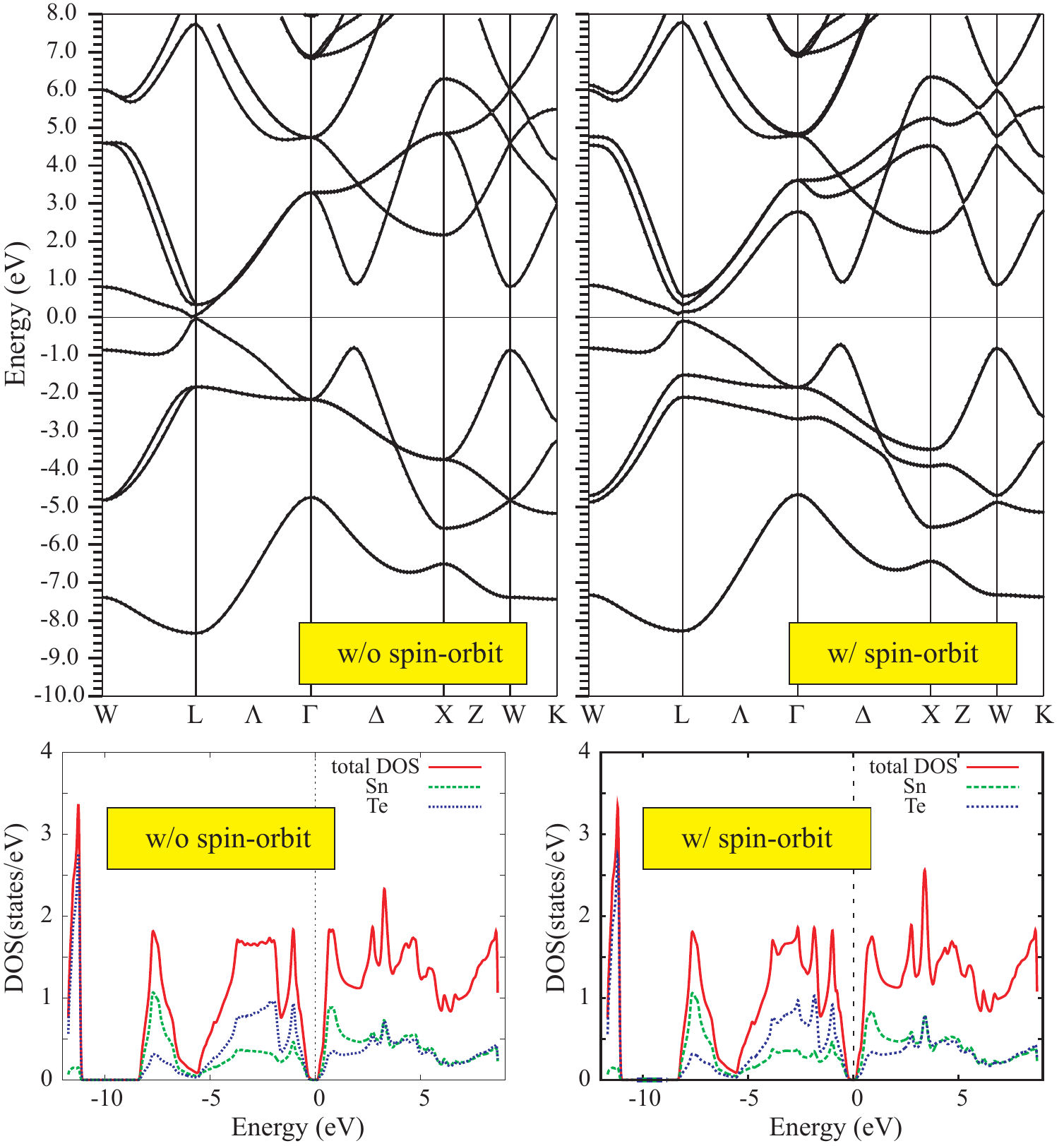}}
\caption{(Color online) Band structure of cubic SnTe calculated without (left panel) and with (right panel) the inclusion of spin-orbit coupling, with the densities of states below. Notice that the pocket near L contributes a small amount to the density of states, and the dominant gap in the spectrum is the roughly 2~eV feature corresponding to the saddle point seen at W and intersected along $\Gamma$-X.}
\label{Fig2}
\end{figure}

Considerable effort over a number of years was invested in the study of the IV-VI band structures~\cite{burke65,tung69,allgaier72,gordyunin74,melvin79}, especially to the complex and spin-orbit-dependent structure near the minimum gap at the L-point. There is general agreement between the results, and our calculations shown in Fig.~\ref{Fig2} are in agreement with the consensus picture. The stoichiometric compound is predicted to be a semiconductor with a narrow gap at the L-point, which lies at the center of the (111) Brillouin zone faces, whereas the average gap is larger ($\sim$2~eV)  elsewhere in the zone.

\begin{figure}[h]
\centering{\includegraphics[width=0.81\columnwidth]{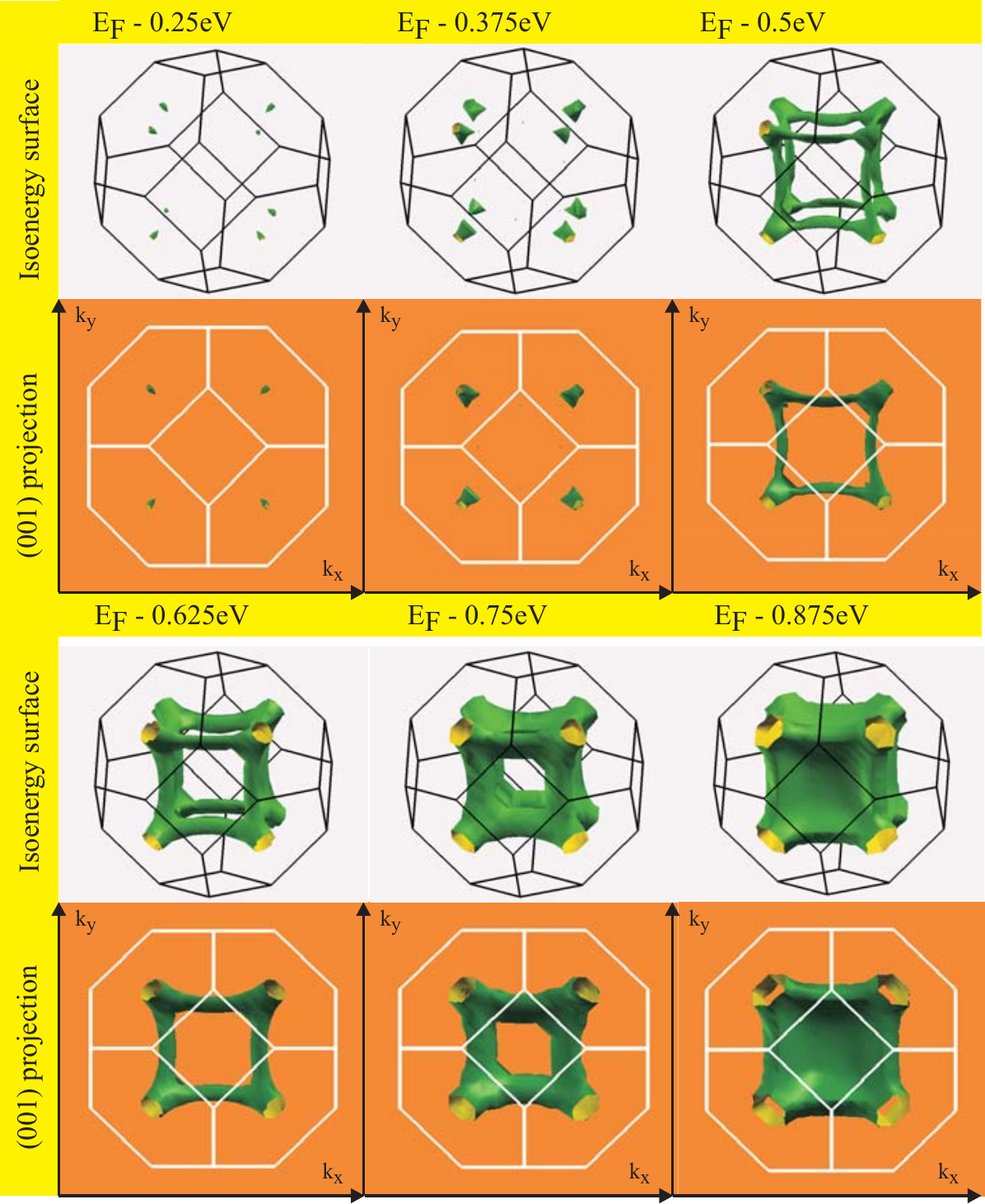}}
\caption{(Color online) Constant energy surfaces from the band structure of Fig.~\ref{Fig2} for a set of different energies. The projected Brillouin zone onto the (001) plane is shown for reference in the lower panels, whereas the upper panels provide a 3D rendering at an oblique angle which makes the topological structure evident.}
\label{Fig3}
\end{figure}

For the purpose of comparison with Fig.~\ref{Fig1},  in Fig.~\ref{Fig3} we show projected energy surfaces at a set of energies below the chemical potential of the undoped system in Fig.~\ref{Fig3}. (Note that the calculated constant-energy $k_x$,$k_y$ contours  illustrated in Fig.~\ref{Fig3} are rotated by 45$^\circ$ with respect to the experimental contours depicted in Fig.~\ref{Fig1}.)
The similarity of these to the data is evident, provided one shifts the chemical potential downward by 0.5~eV. The character of the orbits is perhaps unexpected, but it is made clearer in the 
upper panel for each energy  where we depict the 3D isoenergy surfaces.
We see in Fig.~\ref{Fig3} that the constant energy surfaces evolve from pockets, to interconnected tubes, to sheets. The identification of the experimental Fermi surface as tube-like (rather than sheet-like) is consistent with the absence of any background intensity away from the lines, in contrast to the spectra at -0.5~eV and below.

The evolution of the energy surfaces with hole doping seen in Fig.~\ref{Fig3} was in fact predicted some time ago~\cite{allgaier72} and it is worth revisiting a simple tight-binding model of the bands that contains the salient features, particularly well explicated by Volkov {\em et al.}~\cite{volkov82}. The six bands within a few~eV either side of the chemical potential are predominantly of p-orbital character, with three filled bonding bands and three empty antibonding states. In a cubic structure, including nearest-neighbour hopping $t_\sigma$, $t_\pi$ representing the matrix-elements in the $\sigma$ and $\pi$ coordination, and shifting the atomic orbitals relative to each other by an amount $2 \epsilon$ to account for the ionicity, one has a band structure that accounts for almost all the features of the bands, viz.
\begin{equation}
E = \pm \Bigl \{ \epsilon^2 + 4 \Bigl [ t_\sigma \cos \frac{k_x a}{2}
                                                              + t_\pi \Bigl ( \cos \frac{k_y a}{2}  + \cos \frac{k_z a}{2}  \Bigl ) \Bigl ]^2 \Bigr \}^{\frac{1}{2}}
\>.
\label{pbands}
\end{equation}
This is the dispersion for the pair of $p_x$ bands with those for the $p_y$ and $p_z$ bands obtained by cyclic permutation of the momenta $k_x$, $k_y$, $k_z$ ($a=6.315$\AA\ is the cubic lattice constant). If the ionicity is turned off ($\epsilon=0$) one has three half-filled (metallic) p-bands, mimicking a ``cubic'' group~V material. Eq.~\eqref{pbands} has extrema at the energies $\pm \epsilon$, which occur on surfaces satisfying
\begin{equation}
\cos \frac{k_x a}{2} = - \frac{t_\pi}{t_\sigma} \Bigl ( \cos \frac{k_y a}{2} + \cos \frac{k_z a}{2}  \Bigr )
\>,
\label{extrema}
\end{equation}
(plus cyclic permutations). Because $t_\pi$ and $t_\sigma$ have opposite signs (and relative magnitude approximately 1:5) the six surfaces formed by Eq.~\eqref{extrema} enclose a deformed cube that has corners at the eight L-points of the Brillouin zone, where the surfaces are concave toward the $\Gamma$-point. If one looks for contours of constant energy at energies lower than $-\epsilon \approx 1$~eV, these form two interlocking deformed cubes, of opposite convexity. This is consistent with the energy surfaces seen in the panels of Fig.~\ref{Fig1}, at an energy -1.5~eV.

The approximation of neglecting the coupling between the differently-oriented $p-orbitals$ manifestly fails near the $L$-points, since it predicts a triple degeneracy along the (111) directions, i.e. $\Gamma$-L. This is approximately true for the trio of bands above $E_{\mathrm{F}}$, but not for those below. The L-points are further special, because here the nearest-neighbor hybridization vanishes, and to this order the states are atomic: either Te (lower bands) or Sn (upper bands). In the tight-binding prescription the leading mixing comes from overlap of, for example Sn $p_x$ at the origin with the four neighboring face-centered Sn $p_z$ orbitals in the $x-z$ plane. This leads to a Hamiltonian matrix element
\begin{equation}
\langle A\; p_x | H | A \;p_z \rangle = 4 \, t_{A} \sin\frac{k_x a}{2} \, \sin\frac{k_z a}{2}
\>,
\label{2ndoverlap}
\end{equation}
(and cyclic permutations) that connect amongst the A=Sn or A=Te bases separately, but do not hybridize between the two atoms. The effects of the second shell terms are generally small throughout most of the zone (they vanish on two intersecting planes), but are largest at $L$. At the L-point the Hamiltonian is straightforwardly diagonalized, and the two three-fold degenerate bands are split to make four levels: $-\epsilon-4t_{\mathrm{Te}}$(x2); $-\epsilon + 8t_{\mathrm{Te}}$; $\epsilon-4t_{\mathrm{Sn}}$(x2); $\epsilon+8t_{\mathrm{Sn}}$. Inspection of Fig.~\ref{Fig2} indicates that $t_{\mathrm{Sn}}$ is fairly small and may be neglected, whereas not surprisingly the more occupied Te orbitals overlap strongly. The next matter to note is that if $8 t_{\mathrm{Te}} > 2 \epsilon$, the upper Te orbital crosses the lowest Sn band. This has the effect of pinning the doubly degenerate Sn L-point state to the Fermi level (three bands are filled) so that to this level of approximation the material would be a {\em zero}-gap semiconductor (in the manner of gray Sn). Other weak perturbations in fact produce a small splitting, as do spin-orbit interactions that lower the degeneracies further, leaving the material a narrow gap semiconductor at stoichiometry.

We see that the dispersion near the L-point is anomalous because it is at this high symmetry point that the weak perturbations come together. But the thermodynamics and the structural properties of the IV-VI materials are all explained simply through the large gap 2$\epsilon$, which dominates most of k-space, as was demonstrated before~\cite{pbl80}.

We now have a straightforward explanation of the apparently exotic energy surfaces. Very close to the Fermi energy (and unobserved in the current experiment), one has eight small but highly anisotropic pockets. At slightly lower energies, these pockets trifurcate~\cite{gordyunin74} and join along tubes that are produced by the weak hybridization of the intersecting surfaces from Eq.~\eqref{extrema} owing to Eq.~\eqref{2ndoverlap}. These tubes at their maximum extent enclose only a small percentage of the volume of the Brillouin zone. Then, quite dramatically at an energy that matches the average gap, 2$\epsilon$, the topology changes again into a pair of deformed cubes, when the energy drops below the extrema of the main part of the p-bands.

\begin{figure}[t]
\centering{\includegraphics[width=0.9\columnwidth]{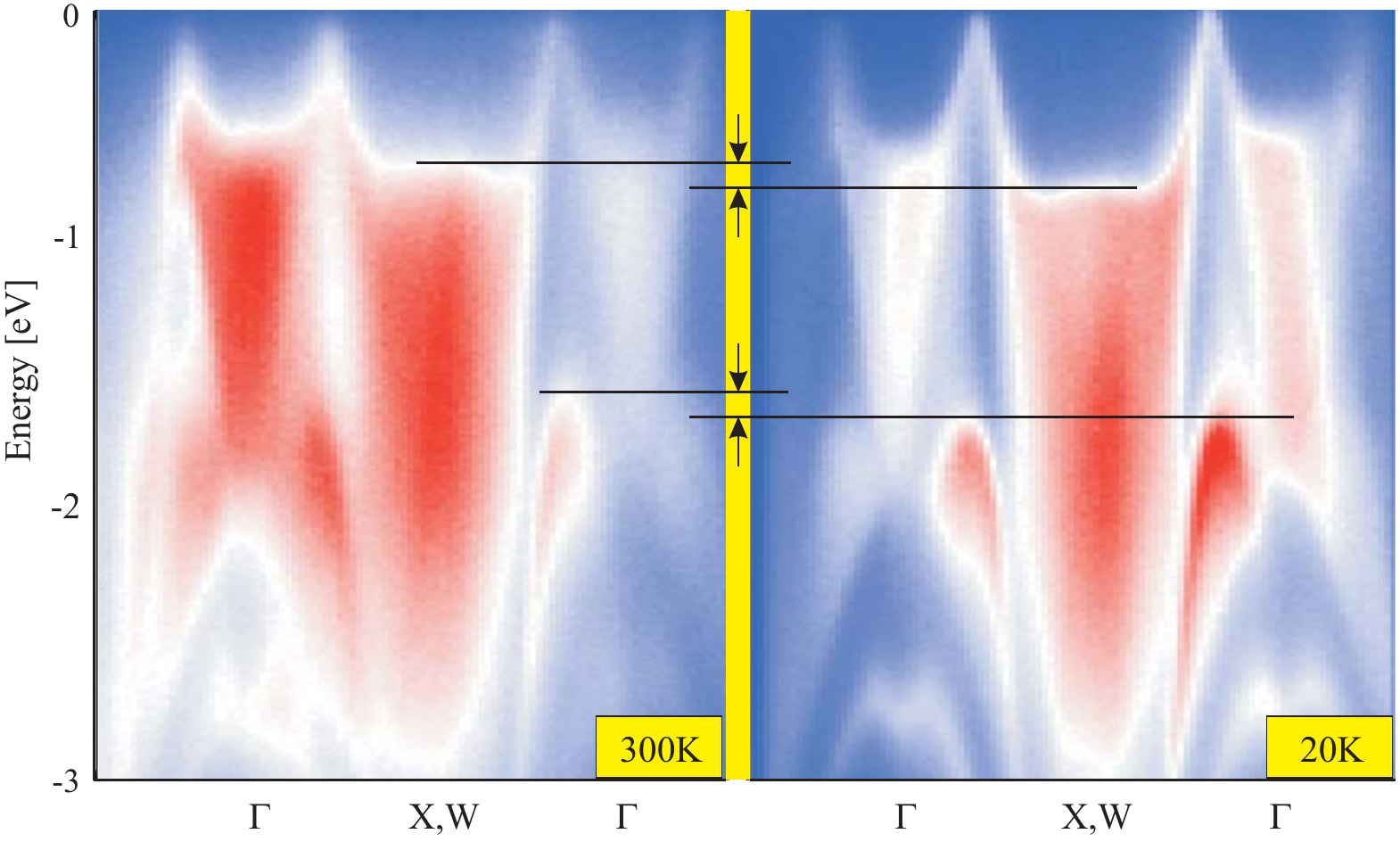}}
\caption{(Color online) Dispersion projected onto the (100) direction at two temperatures. Away from the tube position one sees a weakly dispersive and sharply defined edge at all temperatures, which shifts downward by $\sim 0.1$~eV in the low temperature state. A similar shift is seen in the lower band maximum between $\Gamma$ and $X$ which can be identified from Fig.~\ref{Fig2} }
\label{Fig4}
\end{figure}

Below 100~K SnTe deforms into a rhombohedral structure, driven by a small dimerization in the unit cell, akin to that seen in Sb. 

The driving force for the transition is the lowering in energy that occurs since the large density of states peaks at the energies $\pm \epsilon$ shift to $\pm \sqrt {\epsilon^2 + g^2 u^2}$, where $g$ is the electron-phonon coupling and $u$ the internal relative displacement of the Sn and Te atoms along the (111)-directions. 

The transition is driven by the energy lowering due to the shift of the large peaks in the density of states from energies $\pm \epsilon$ to $\pm \sqrt {\epsilon^2 + g^2 u^2}$. Here, $g$ is the electron-phonon coupling and $u$ the internal relative displacement of the Sn and Te atoms along the (111)-directions. 

Using the internal displacement $\vec u = \pm \frac{1}{2} \tau a (1,1,1)$ with $\tau = 0.008$~\cite{iizumi75} and the coupling constant estimated from pseudopotential theory~\cite{pbl79,pbl80} one predicts a shift by approximately 0.15~eV. 
Searching for this is complicated by the broadening of the spectra at high temperatures, and also since the prominent L-point features near the Fermi energy are weakly affected. However, by comparing spectra that lie off the symmetry points, shifts are revealed, and are shown in Fig.~\ref{Fig4}.

The photoemission results presented here resolve long-standing issues about the nature of carrier transport in heavily doped SnTe, demonstrating that over a range of energies the hole pockets merge to an open network of tubes. We find however, that once the band filling exceeds a few percent, the band structure simplifies to that of three quasi-1D p-bands, which are responsible for the thermodynamic stability of the material. We are also able to resolve the extra gapping of these p-bands which is responsible for the structural transition into a low-temperature polar phase.

This work was performed in part under the auspices of the United States Department of Energy.  PBL thanks Los Alamos National Laboratory for hospitality in the course of this research, also supported by the Engineering and Physical Sciences Research Council, UK.

\bibliography{apssamp}

\end{document}